\documentclass[lnbip]{svmultln}
\usepackage{makeidx}  
\usepackage[scaled=0.86]{helvet}
\usepackage{xcolor}
\usepackage{hyperref}
\usepackage{longtable,booktabs,array}
\usepackage{colortbl}
\usepackage{calc} 
\usepackage{etoolbox}
\usepackage{graphicx} 
\usepackage{todonotes}
\usepackage{doi}
\usepackage{etoolbox}
\usepackage{todonotes}
\usepackage{caption}
\AtBeginEnvironment{longtable}{\scriptsize}

\begin{document}
\mainmatter              
\title{Using Process Mining with Pre- and Post-Intervention Analysis to Improve Digital Service Delivery: A Governmental Case Study}
\titlerunning{Using Process Mining to Improve Digital Service Delivery}  
%
\author{Jacques Trottier\inst{1} \and William Van Woensel\inst{2} \and 
Xiaoyang Wang\inst{3} \and Kavya Mallur\inst{1} \and Najah El-Gharib\inst{3} \and Daniel Amyot\inst{3}}
\authorrunning{Jacques Trottier et al.}   
%
\tocauthor{Jacques Trottier, William Van Woensel,
Xiaoyang Wang, Kavya Mallur, Najah El-Gharib, Daniel Amyot}
\institute{Government of Canada, Ottawa, Canada,\\
\email{jacques@trottier.us, kavyaa.jnr@gmail.com},
\and
Telfer School of Management, University of Ottawa, Ottawa, Canada\\
\email{wvanwoen@uottawa.ca}
\and
School of EECS, University of Ottawa, Ottawa, Canada\\
\email{xwang233@uottawa.ca, nelgh031@uottawa.ca, damyot@uottawa.ca}
}

\maketitle 

\begin{abstract}        
We present a case study of Process Mining (PM) for personnel security screening in the Canadian government. 
We consider customer (process time) and organizational (cost) perspectives. Furthermore, in contrast to most published case studies, we assess the full process improvement lifecycle: pre-intervention analyses pointed out initial bottlenecks, and post-intervention analyses identified the intervention impact and remaining areas for improvement. 
Using PM techniques, we identified frequent exceptional scenarios (e.g., applications requiring amendment), time-intensive loops (e.g., employees forgetting tasks), and resource allocation issues (e.g., involvement of non-security personnel). Subsequent process improvement interventions, implemented using a flexible low-code digital platform, 
reduced security briefing times from around 7 days to 46 hours, and
overall process time from around 31 days to 26 days, on average. From a cost perspective, the involvement of hiring managers and security screening officers was significantly reduced. These results demonstrate how PM can become part of a broader digital transformation framework to improve public service delivery. The success of these interventions motivated subsequent government PM projects, and inspired a PM methodology, currently under development, for use in large organizational contexts such as governments.
\end{abstract}

\section{Introduction}\label{introduction}

Recent issues with passport processing~\cite{Tasker2024} and immigration~\cite{Osman2023} highlight challenges that the Government of Canada (GC) is facing in implementing efficient business processes for delivering services to Canadians. Underpinning much of the GC's current issues with service delivery is the shift to digital services. This shift requires the transformation of aging IT systems, some as old as 50 years~\cite{TBS2022}; in addition, the increasing need for digital service delivery has required governments worldwide to redesign processes that were originally geared towards pre-digital paper-based services. To support such a process redesign, Process Mining (PM) offers evidence-based and data-driven technology to map current business processes, study their performance, diagnose their potential issues, and derive solutions in terms of process improvements. 

In this paper, we demonstrate the utility of PM to improve government services by mining the Personnel Security Screening (PSS) process~\cite{TBS2014} of a particular GC department (see Appendix A.1). 
As this process is a mandatory part in the hiring and vetting of all prospective employees, the GC processes tens of thousands of PSS transactions per year. Hence, while individual departments can choose how to implement the process, an optimized process implementation may provide insights for improving cost and hiring time efficiency across the GC.

The objectives of this case study are to (a)~discover and evaluate process and handover-of-work models for PSS, analyzing process performance and bottlenecks; (b)~present subsequent opportunities for process improvement; and (c)~assess the results from PSS process improvements that followed from opportunities identified in (b).
By using PM techniques, we were able to gain insights into the frequency of problematic scenarios (e.g., applications requiring amendment), time-intensive loops (e.g., employees forgetting tasks), and resource allocation issues (e.g., involvement of non-security personnel).
Leveraging a flexible low-code digital platform, we were able to quickly guide multiple interventions to resolve these issues. 
We found that these interventions improved service delivery from customer (process time) and organizational (cost) standpoints: the part targeted by the interventions, namely security briefings, had its average process times reduced from about 7 days to 46 hours; overall average process time was thus reduced from about 31 days to 26 days. Moreover, involvement of the hiring manager and security screening officer were significantly reduced.

Contributions of this paper include: (1)~the mining of a security clearance process in the GC, with a discussion of lessons learned (during this and other case studies) on conducting PM in a government context; (2)~an empirical analysis of the full process improvement lifecycle, i.e., covering both pre- and post-intervention; 
and (3)~re-usable code for event log cleaning and transformations used in the project, which illustrates the Python library for \textit{Process Mining -- Log Filtering \& Preprocessing} (\texttt{logprep4pm}) developed by the authors~\cite{ElGharib-API-2022,logprep4pm-2024}.

In this paper, Section~\ref{related-work} covers related work. Section~\ref{methodology} elaborates on our methodology, whose results are presented in Section~\ref{results} and further discussed in Section~\ref{discussion}. Section~\ref{limitations} highlights limitations, and Section~\ref{conclusions-and-future-work} provides conclusions.

\section{Related Work}\label{related-work}

The importance of process mining in government services and digital transformation was acknowledged in a recent literature review from Rawiro et al.~\cite{Rawiro2022}, which covered 25 papers between 2009 and 2022 that spanned 18 different countries.
The papers reported case studies ranging from car registration to civil status management, procurement, and fine management, typically covering process discovery (n=14). 
We note that none of the case studies covered GC services or security screening.
The paper closest to ours was the assessment of a passport application process in Uruguay by Delgado et al.~\cite{Delgado2024,Delgado2020}, who introduced a framework for selecting and analyzing target processes, with a focus on process discovery and improvement opportunities. The framework also includes a metamodel for data integration. The authors reported the access to data, quality of event logs, and inter-organizational processes as major challenges.
We also note that recent work started integrating AI into government process mining projects. For example, Nai et al.~\cite{Nai2023} recently used natural language processing for the generation and enrichment of event logs in support of the discovery and analysis of procurement processes from France, Spain, and Italy.

To our knowledge, outside of healthcare, there is a lack of case studies that cover the full process improvement lifecycle---i.e., before and after specific interventions.
Zuidema-Tempel et al.~\cite{Zuidema-Tempel2023} compare PM methodologies with PM practitioner experiences;
they report that while post-analysis and monitoring in industry do happen in practice, PM methodologies lack focus on quantifying, selecting and \textit{monitoring} improvement actions. Leemans et al.~\cite{Leemans2019} focus on process compliance and performance at a Queensland Government department, but do not address actual process improvements. 
Park et al.~\cite{Park2016} compared process mining vs. traditional process re-engineering 
in government; they shortly discuss process changes before and after interventions, but mainly focus on differences in processing times between municipalities. 

\section{Methodology}\label{methodology}
During the COVID-19 pandemic, a GC department\footnote{Due to issues of confidentiality, we cannot identify the actual department.} implemented a new PSS system, which includes an online web-based portal for new/prospective employees to complete their security screening application. Its goal was to facilitate a more contactless experience while improving processing times. The PSS system was built on an industry-recognized, low-code Business Process Management (BPM) platform, which 
records events throughout a case's lifecycle. Moreover, as mentioned, the GC processes tens of thousands of PSS applications per year; an optimized process could pave the way for large gains in efficiency throughout the GC. These factors made it a highly suitable candidate for PM.

Our goal was to improve the \textit{security briefing} part of PSS.
Our initial research question thus revolved around finding bottlenecks from two perspectives:

\vspace{-0.20cm}
\begin{itemize}
    \item \textit{Process time}: time for an applicant to complete the security briefing.
    \item \textit{Cost}: salary implications of the GC employees involved in the briefing sessions.
\end{itemize}
\vspace{-0.15cm}

Using PM, we were able to answer this question, which informed improvement interventions that were then implemented.
Subsequently, our research question was refined to measure the interventions' impact on process time and cost. 

The project included 3 distinct phases described in Table~\ref{tab:Phases}. For each phase, we applied the methodology from Table~\ref{tab:Steps}. Below, we discuss general aspects that apply to all phases.
For the \emph{extraction of raw event logs} step, the logs were extracted from the SQL database of the PSS system. For pre-processing and event log generation, we used JupyterLab with data analysis libraries, including a Python port (\texttt{logprep4pm})~\cite{logprep4pm-2024} of the \emph{Cloud Pattern API for Process Mining}~\cite{ElGharib-API-2022}.

\begin{longtable}[]{@{}
  >{\raggedright\arraybackslash}p{(\columnwidth - 4\tabcolsep) * \real{0.10}}
  >{\raggedright\arraybackslash}p{(\columnwidth - 4\tabcolsep) * \real{0.25}}
  >{\raggedright\arraybackslash}p{(\columnwidth - 4\tabcolsep) * \real{0.65}}@{}}
\caption{Project phases, date ranges of analyzed event data, and activities}
\label{tab:Phases}\\
\toprule()
\begin{minipage}[b]{\linewidth}\raggedright
\textbf{Phase}
\end{minipage} & \begin{minipage}[b]{\linewidth}\raggedright
\textbf{Event Date Range}
\end{minipage} & \begin{minipage}[b]{\linewidth}\raggedright
\textbf{Activities}
\end{minipage} \\
\midrule()
\endhead
I & \begin{minipage}[t]{\linewidth}\raggedright
Jan 6 -- Oct 26, 2022\strut
\end{minipage} & Initial Analysis (Process Discovery \& Social
Mining) \\
\rowcolor{gray!10}II & \begin{minipage}[t]{\linewidth}\raggedright
Jan 5 -- May 15, 2023\strut
\end{minipage} & \begin{minipage}[t]{\linewidth}\raggedright
Intervention I (Group Briefing) \& Subsequent Analysis\strut
\end{minipage} \\
III & Jun 15 -- Oct 23, 2023 & Intervention II (Video Briefing) \& Subsequent Analysis \\
\bottomrule()
\end{longtable}

\begin{longtable}[]{@{}
  >{\raggedright\arraybackslash}p{(\columnwidth - 6\tabcolsep) * \real{0.2597}}
  >{\raggedright\arraybackslash}p{(\columnwidth - 6\tabcolsep) * \real{0.2744}}
  >{\raggedright\arraybackslash}p{(\columnwidth - 6\tabcolsep) * \real{0.2078}}
  >{\raggedright\arraybackslash}p{(\columnwidth - 6\tabcolsep) * \real{0.2581}}@{}}
\caption{Steps for the 3 phases. Steps with (*) were only conducted in Phase I.}
\label{tab:Steps}\\
\toprule()
\begin{minipage}[b]{\linewidth}\raggedright
\textbf{Activity}
\end{minipage} & \begin{minipage}[b]{\linewidth}\raggedright
\textbf{Chosen Tools}
\end{minipage} & \begin{minipage}[b]{\linewidth}\raggedright
\textbf{Result / Output}
\end{minipage} & \begin{minipage}[b]{\linewidth}\raggedright
\textbf{Reason}
\end{minipage} \\
\midrule()
\endhead
Extraction of raw event logs & MS SQL Server Management Studio 18 & Raw
Event Log (CSV) & Compatibility with backend DB (MS SQL) \\
\rowcolor{gray!10}Pre-processing of the event log & 
\texttt{logprep4pm}~\cite{logprep4pm-2024} library
Pandas library

JupyterLab & \begin{minipage}[t]{\linewidth}\raggedright
Preprocessed Event Log (CSV)

.ipynb notebook\strut
\end{minipage} & \begin{minipage}[t]{\linewidth}\raggedright
\texttt{logprep4pm} and Pandas to clean and filter event logs with re-usable code\\

JupyterLab for documenting steps\strut
\end{minipage} \\
\begin{minipage}[t]{\linewidth}\raggedright
Process model\\
discovery\strut
\end{minipage} & Disco & \begin{minipage}[t]{\linewidth}\raggedright
Directly\\
Follows Graph\strut
\end{minipage} & Discover a process model from the event log data \\
\rowcolor{gray!10}Exporting XES event log (*) & Disco & XES file & ProM requires XES
file \\
\begin{minipage}[t]{\linewidth}\raggedright
Social Network Model\\
Discovery (*)\strut
\end{minipage} & ProM & Handover-of-work model & Disco does not support
social network mining \\
\bottomrule()
\end{longtable}

The \texttt{logprep4pm} library overlays filtering and pre-processing functionalities for PM on top of the popular Pandas library. An important benefit of 
using a code library for this purpose is code reuse: 
code written for complex (yet common) exploratory data analysis, filtering, and pre-processing, can be re-used across phases, and shared within the organization (e.g., GC), regardless of PM tools used.
We provide links to abstracted and anonymized notebooks that document these exploratory data analysis, filtering, and event log pre-processing steps using the \texttt{logprep4pm} library, among other libraries, in~\hyperref[subsec:appendix-artifacts]{Appendix A.4}.


During the \emph{pre-processing of event logs}, we observed 8 event classes (i.e., types of events) for which there were less than 10 occurrences; these were deleted to prevent them from cluttering the process. Also, we observed duplicate events in short succession; we attributed this to users accidentally double clicking a button. 
\texttt{logprep4pm} provides a function to delete duplicate events within a time threshold; 
we selected a time threshold of 3 minutes (i.e., events separated by 3 minutes or more were assumed to be separate actions). This seemed appropriate based on the event data and nature of the tasks, and removed most of the duplicate events (10 in total). Using \texttt{logprep4pm}, case ID numbers were also anonymized to protect privacy, and event names were renamed to user-friendly descriptions. Employee emails were replaced with the corresponding role.

For each phase, in line with the PM$^2$ methodology~\cite{PM2}, we applied several iterations of \emph{pre-processing, mining \& analysis,} and \emph{evaluation} to refine the event data and eliminate data quality issues. Throughout these cycles, our event log was exported multiple times as a CSV file from \texttt{logprep4pm} (Jupyter Notebook) and imported into the Disco PM tool. Iterations continued until we were satisfied that the discovered model accurately represented the business process with minimal noise, in collaboration with the process owner (i.e., GC department).


\subsection{Phase I -- Initial Process Discovery \& Social Network Mining}
\label{phase-i-initial-process-discovery-social-mining}

The event logs generated shortly after the PSS launch in April of 2021 were noisy. 
Given the sensitive nature of PSS for higher-level clearances, this case study's scope was also limited to exploring security briefings for ``reliability status'', the lowest security clearance level. After filtering, this resulted in an initial case count of 273. 
We used the methodology from Table~\ref{tab:Steps} to answer our initial research question on process time and cost for the PSS process, i.e., by discovering and studying the associated process and social network models. 

\subsection{Phases II and III -- Interventions}
\label{phase-ii-and-iii-interventions}

Two interventions were applied to the PSS process to improve processing times. 
We note that the original process included a one-on-one security briefing between the new employee and hiring manager (see~\hyperref[subsec:appendix-pss-process]{Appendix A.1}). The \textbf{first intervention} replaced these one-on-one briefings with a group briefing session with the security screening officer. These group sessions were scheduled 4 times per week. 
This intervention took place at the end of October 2022, and the resulting process lasted until June 2023 (when the second intervention took place). We extracted 191 cases from January 5 to May 15, 2023; as before, we chose these times to cope with bug fixes and process drift right after and before interventions.

The \textbf{second intervention} subsequently replaced these group briefings with a non-proctored video that applicants would watch by logging into the PSS web portal. Applicants would digitally attest to having watched the video and accept the conditions of their clearance. This is the solution that is still in place at the time of writing (August 2024). The intervention took place in June 2023. We extracted 182 cases during this time period, from June 15 (considering a buffer after implementation) until October 23, 2023 (time of analysis). Log inspections confirmed that there was no overlap from different phases.

We discuss how these interventions were motivated by PM result analyses in Section~\ref{results}.
After each intervention, we refined our research question to measure the resulting process time and cost, which involved re-applying our methodology.


\section{Results}\label{results}

\subsection{Phase I - Initial Process Discovery \& Social Network Mining}
\label{phase-i---initial-process-discovery-social-mining}

\begin{figure}
    \centering
    \includegraphics[width=0.7\linewidth]{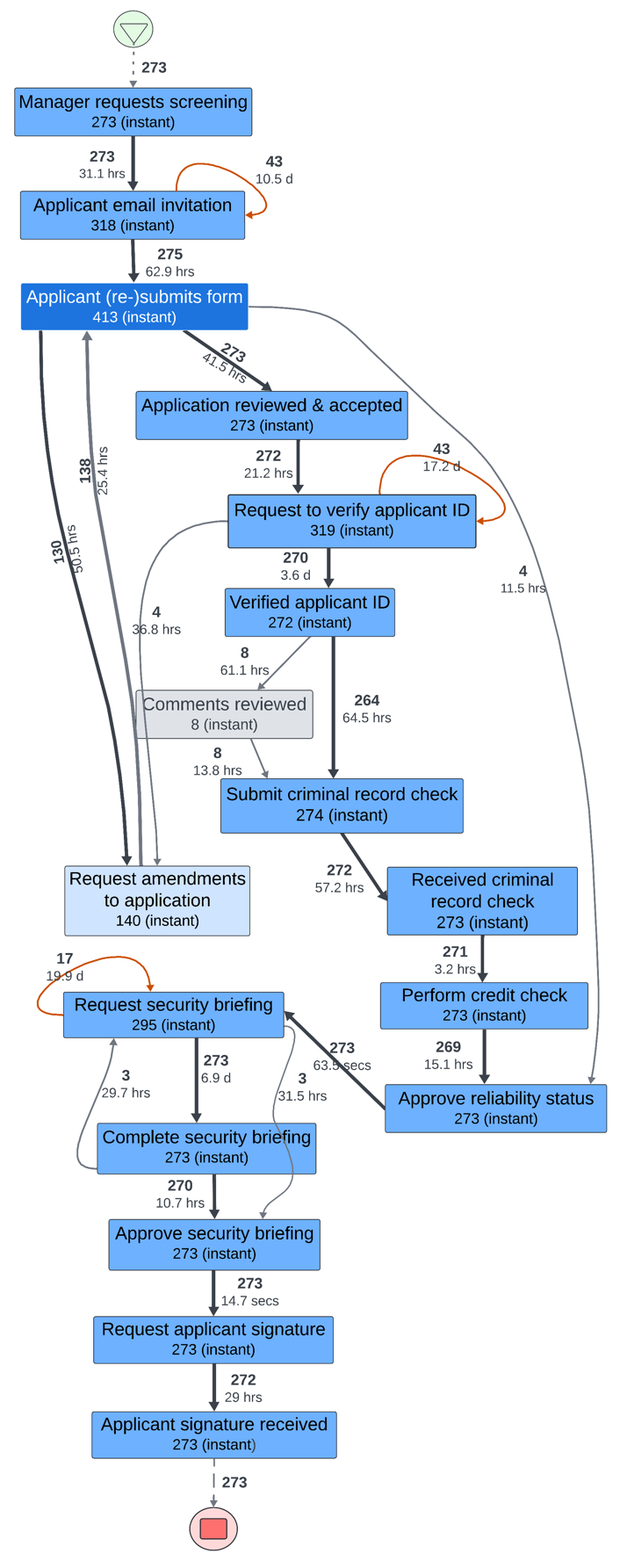}
    \caption{Process model discovered using Disco and reformatted (Phase I). Links with two or fewer cases were left out for clarity. Exceptional activity flows mostly cover cases where other steps (e.g., ``Applicant (re-)submits form'' to ``Approve reliability status'') had already been performed. Numbers in bold indicate frequencies; numbers below them indicate durations.}
    \label{fig:ProcessModel}
\end{figure}

The initial process model is illustrated in Fig.~\ref{fig:ProcessModel}. Mean and median case durations were 30.9 days and 19.9 days, respectively (this included weekends, vacations, and public holidays). We found that this process model generally complies with the standard for security screening~\cite{TBS2014}.
We summarize here 3 major findings.

\subsubsection{Finding 1 -- Data Validation.}
\label{finding-1-data-validation}

We found that scenarios involving security screening applications required amendments very frequently. In particular, 48\% (130/273; Fig.~\ref{fig:ProcessModel}) of security screening applications were refused after initial submission, mainly because of missing/inverted names and outdated ID issues. 
We list the main reasons why an application required amendments in \hyperref[subsec:appendix-a3]{Appendix A.3}.
Several solutions were recommended to mitigate this issue. In the short term, the online application form was extended with tooltips (or hints) in the ``full given names'' field to encourage applicants to include their middle name and exclude their surname. In the medium term, name, address, and date extraction from uploaded ID documents (using optical character recognition -- OCR) was recommended, while in the long term a better integration with federal and provincial systems would help pre-populate that information from reliable sources.


\subsubsection{Finding 2 -- Event Loops.}\label{finding-2-event-loops}

We found multiple time-wasting loops in the initial process model (Fig.~\ref{fig:ProcessModel}):

\begin{enumerate}
\def\labelenumi{\arabic{enumi}.}
\item
  \emph{Applicant Email Invitation:} 
  An email is sent to the applicant requesting they register to the portal and complete the
  application form. In 43 instances, an email invitation had to be re-sent as the applicant had not registered yet (e.g., missed the email, went to junk folder, lack of digital competency).
\item
  \emph{Request to Verify Applicant ID:} 
  The hiring manager must verify the ID documents from the application submission. 
  In 43 instances, the hiring manager had to be reminded to conduct this activity. We suspect this was due to the managers' busy workload and large volume of emails.
\item
  \emph{Requested Security Briefing:} 
  An approved applicant must be briefed on their security responsibilities by the hiring manager. In 17 instances, this briefing activity had to be re-requested (likely for similar reasons as above).
\end{enumerate}

To partially address these issues, it was suggested that the system could automatically send daily reminders to the applicant and hiring managers to complete outstanding tasks. Some reminders were implemented after our study.

\subsubsection{Finding 3 -- Social
Mining.}\label{finding-3-social-mining}

A handover-of-work model, as an output of social network mining, captures a relation between stakeholders \textit{X} and \textit{Y} when, for two sequential activities \textit{A1} and \textit{A2}, \textit{X} does \textit{A1} and \textit{Y} does \textit{A2}. 
This model also captures the size and influence that each stakeholder plays in a given process. ~\hyperref[subsec:appendix-how]{Appendix A.2} contains the handover-of-work model as generated by ProM.

We found that the \emph{Applicant} had the largest participation in the process, and the \emph{Credit Bureau} had the least. Rather surprisingly, however, security screening resources---including \emph{Security Officers} and \emph{Security Systems}, which should be pivotal in the process---were only involved in 28\% of the process lifecycle, while the \textit{Hiring Manager} was involved to a similar degree (26\%), although security is not their primary role.
We further observe that \textit{Security Officers} acted mainly as facilitators as  
all other stakeholders but one (\textit{Security Systems}) involved handover work to the security officer. 
Once all necessary information had been gathered, they rendered a decision of whether to grant reliability status. 
This process put a large strain on \emph{Hiring Managers}; we found that the security briefings (which last 10-15 minutes) were a major contributor to their involvement.
A separate briefing had to be scheduled per \emph{Applicant}, which delayed the process by 7 days on average (Fig.~\ref{fig:ProcessModel}). 
We had suspected (see prior section) that their busy workload had led to process delays and time-wasting loops.

In line with this observation, two interventions
were proposed by the process owner to re-organize these security briefings. 
Our research question was thus refined to measure process time and cost of post-intervention security briefings.


\subsection{Phase II: Group Briefing Intervention}
\label{phase-ii-group-briefing-intervention}

The Group Briefing intervention involved 1)~transferring the \emph{Hiring Manager's} briefing-related duties to the \emph{Security Officer} (in particular, the Security Screening Officer, SSO), and 2)~moving from one-on-one briefings to group-based briefings organized at fixed times each week.

\begin{figure}
    \centering
    \includegraphics[width=0.85\textwidth]{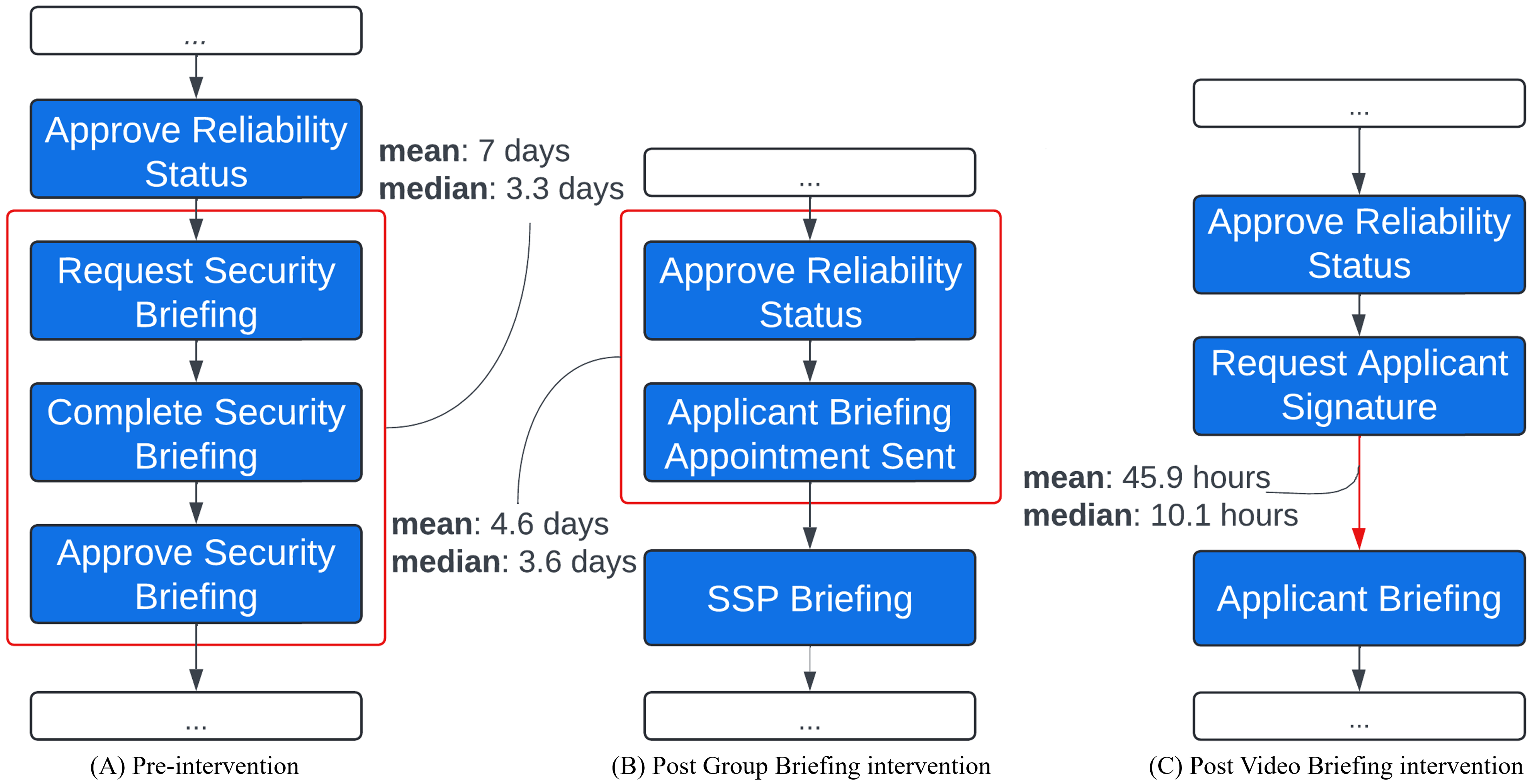}
    \caption{Process models pertaining to security briefings.}
    \label{fig:SecurityBriefings}
\end{figure}

Figure~\ref{fig:SecurityBriefings} zooms in on part of the process model that pertains to security briefings: (A) shows the section from the original model, 
whereas (B) and (C) show the post-intervention sections, i.e., after the Group Briefing and Video Briefing interventions.
The parts highlighted in red are specifically related to requesting and completing the security briefing.
Regarding (B), i.e., Group Briefing intervention, we can answer our research question as follows:

\begin{itemize}
\item
  \emph{Process time}: the process enters the relevant stage in (A) when the hiring manager is supposed to schedule a briefing session with the applicant; in (B), when the applicant receives an automated request to register for a group briefing. We point out that this time includes service time as well as waiting for the applicant. The mean process time was reduced from 7 to 4.6 days; the median process time however stayed more or less the same.
\item
  \emph{Cost}: moving to group-based briefings meant a reduction from 191 sessions (1 per case) to 76 sessions (4 per week) over the time period, leading to a 60\% reduction in number of sessions. Conservatively, assuming 10 minutes per briefing, this saves 
  about 18 hours of organizational time and associated salary over the time period.
  The government department performs around 3000 clearances per year (including secret and top-secret instances, not discussed in this case study); avoiding one-on-one briefings thus saves around 500 hours of hiring manager time annually.
\end{itemize}

\subsection{Phase III: Video Briefing Intervention}
\label{phase-iii-video-briefing-intervention}

The Video Briefing Intervention, which was developed concurrently with the first intervention, replaced the Group Briefing with a non-proctored online briefing video. Hence, the intervention removed employee involvement altogether, and allowed applicants to perform the security briefing at their convenience. Figure~\ref{fig:SecurityBriefings}~(C) shows the result of this new intervention.
Regarding our research question on process times and cost:
\begin{itemize}
\item
  \emph{Process time}: the process enters this stage when the applicant receives an automated request for
  watching and digitally attesting to the online video. 
  As before, this time includes both service time and waiting for the applicant. The mean process time was reduced from 4.6 days to 45.9 hours; the median process time was reduced from 3.6 days to 10.1 hours. Comparing Phases I and III yields an overall time savings of around 5 days on average.
\item
  \emph{Cost}: as there is no more involvement required from GC employees during briefings, this removed any employee time involvement and associated salary implications. Avoiding group briefings thus saved around 35 hours of SSO time annually (4 briefings for 52 weeks). The initial cost of recording the video, and updating the process on the BPM platform, was quickly amortized over time. Cumulatively, the impact of these two interventions led to approximately 535 hours of departmental time savings annually.
\end{itemize}

\section{Discussion and Lessons Learned}\label{discussion}

We make the following observations:

\textit{Process mining was instrumental in the problem identification}. We found frequent erroneous sequences, time-wasting loops, long delays, and unexpected degrees of participation, in the PM models. 
While we focused on relatively basic PM techniques, they were nevertheless effective at identifying these problems. A bonus here is that the algorithms and their output are easy to understand, seeing how it was the first time that PM was applied in this government department.

\textit{PM showed a clear potential for a Return On Investment (ROI)}.
The interventions 
could be seen as obvious improvements that did not necessarily require insights from PM results.
In an organizational setting, however, even straightforward improvements will involve resources for their implementation, and we found that they require a clear indication of their ROI (we revisit this issue in our conclusion). To that end, our pre-intervention analysis pointed out that process time (mean 7 days for a 10-15 minute briefing) and cost (Hiring Managers were overly involved) were problematic; and, post-intervention analysis for the Group Briefing confirmed the need for the Video Briefing intervention, as process times (similar median, 4.6 days mean) were still high. 

\textit{Process time and cost were significantly reduced after the final intervention}. 
The Group Briefing reduced the mean process times by around 2.5 days (no impact on median); the Video Briefing intervention, compared to the initial process, reduced mean process times by about 5 days (median savings of about 3 days). Regarding cost, the Group Briefing reduced the number of briefing sessions by around 60\%, saving around 500 hours of hiring manager time annually; the Video Briefing intervention removed employee involvement altogether. 

\textit{Lessons learned from PM in a governmental context.}
We performed this case study as a centralized team of technology experts, called a ``Centre of Excellence'' (CoE), to collaborate with the government department. 
This setup is common in large organizations for a relatively novel and challenging technology. However, this context gave rise to, or at least aggrevated, a number of issues. These issues were not technical, but involved the need for (a)~\textit{process selection}, i.e., prioritizing candidate processes and readiness  from departments for managing risk and maximizing value; (b)~\textit{establishing value (pre- and post)}, i.e., engaging process owners a priori by presenting an ROI for resource allocation, and then measuring concrete impacts after interventions; and (c)~\textit{domain understanding}, i.e., continuous communication with process owners to understand the process and interpret analysis findings. We revisit this in the future work.

\textit{Findings from the study can be applied to other GC departments}. The results applied to a single department. The GC employs over 300,000 individuals and has over 100 departments, agencies and crown corporations that implement their own version of the PSS process; applying the findings from this case study across the GC could thus yield substantial cost and time benefits. Depending on the similarities of their PSS process, findings could be directly used to improve their own process;
otherwise, a methodology similar to the one presented in Section~\ref{methodology} could be used 
to guide potential and customized interventions.


\section{Limitations}\label{limitations}

Limitations of our study include the following four items.

\emph{Restricted number of cases used in the study}. The sample cases (around 190 for each intervention) is only a portion of the total number of security clearance cases (3000) in the department, which is itself a fraction of the cases in the GC. The process model and performance metrics may not fully capture process variations in the entire GC, as each GC department can implement its own bespoke version of the Treasury Board of Canada's standard for security screening.

\emph{Potential seasonal process variations}. Different time periods may have associated differences in workload that could impact durations (e.g., process times) of security screenings. Due to time constraints, we were unable to select cases from the same annual period for the group briefings (Jan 5, 2023, to May 15, 2023) and video briefings (June 15, 2023, to Oct 23, 2023, our study endpoint).

\emph{Lack of event durations}. Only the start time of each event is recorded;
the event duration (service time) is thus calculated as the difference between the current and next event's start times. Hence, we are unable to distinguish between the service time of an activity and the wait time to reach the next activity.

\emph{Limited to reliability status screening.} These findings apply to reliability status screenings only; outcomes might differ for higher level clearances.

\section{Conclusions and Future Work}
\label{conclusions-and-future-work}

We presented a case study of PM in government, which focused on optimizing personnel security screening from customer (process time) and organizational (cost) perspectives. We used PM to assess the full process improvement lifecycle: pre-intervention analyses pointed out initial process and performance issues; post-intervention identified the impact of interventions and further areas for improvement. Due to the use of a low-code, process-aware BPM platform, the turnaround time for interventions was relatively low. 
The applied interventions successfully reduced the process time from about 30.9 days down to 25.8 days on average, and removed employee involvement and associated salary implications.

The lessons learned from this case study (Section~\ref{discussion}), and subsequent studies we have conducted in the meantime, will be used a basis for a PM methodology for use in large governmental contexts.
Compared to our lessons learned (Section~\ref{discussion}),
PM$^2$~\cite{PM2} similarly considers PM projects from a CoE viewpoint, and discusses process selection (a) and domain understanding (c). 
We further found establishing value (b) to be essential. The proposed PM methodology will further flesh out these aspects.

From the GC perspective, 
its Policy on Service and Digital~\cite{TBS2020} emphasizes that digital transformation is ultimately focused on enhancing the customer experience. We argue that PM will play an important role in supporting this policy, by ensuring that public services are delivered in a timely and cost-effective manner, in compliance with service-level agreements. Finally, by illuminating often opaque government processes in an evidence-based way, PM has the potential to support the Open Government goals of greater openness and accountability.

\subsubsection*{Acknowledgements.}
This work was partially supported by GC's Innovation Funding and by the NSERC Discovery grants of W. Van Woensel and D. Amyot. We thank the process owners and department managers for their collaboration.
%
%
%
\bibliographystyle{splncs04}
\bibliography{references-arxiv}
\section*{Appendix A: Supplemental Material}
\label{sec:appendix}
This appendix provides supplemental material for the paper ``Using Process Mining with Pre- and
Post-Intervention Analysis to Improve Digital Service Delivery: A Governmental Case Study''.

\subsection*{A.1 Personnel Security Screening Process}
\label{subsec:appendix-pss-process}

The Government of Canada processes tens of thousands of Personnel Security Screening (PSS) transactions per year. As a condition of employment, public servants working for the GC must undergo a security screening process. This rule also applies to contractors and third parties working with the GC.

During the COVID-19 pandemic, a GC department implemented a new PSS system, including an online web-based portal for new/prospective employees to complete their security screening application. The goals of this platform were to 1)~improve processing times, for instance, by reducing data validation errors; and 2)~facilitate a more contactless experience due to the COVID-19 pandemic. This new system was a highly suitable candidate for PM as it was built on an industry-recognized, low-code Business Process Management (BPM) platform. As a Process-Aware Information System (PAIS), the platform records key events throughout a case's lifecycle at an adequate level of abstraction. Below, we summarize the PSS, its high-level process, and how the PSS system fits in.

\begin{figure}
    \centering
    \includegraphics[width=\textwidth]{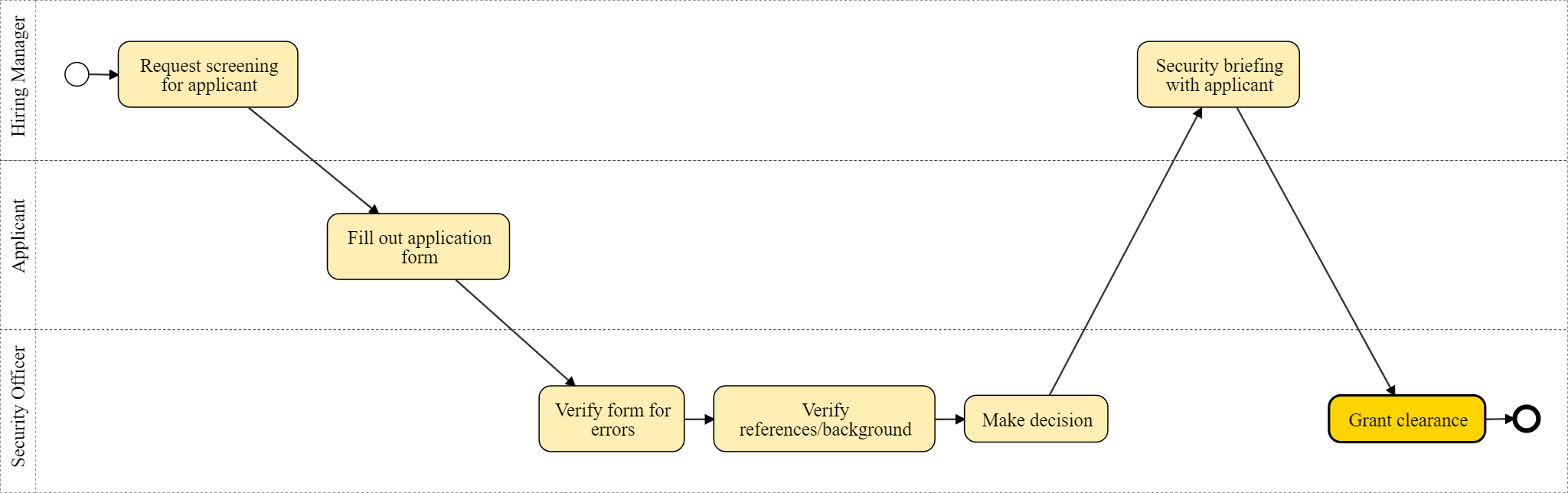}
    \caption{Personnel Security Screening (PSS) normative process overview}
    \label{fig:PSS-process}
\end{figure}

Different security levels exist, starting from \emph{reliability status} and ending with \emph{secret} and \emph{top-secret} clearances. The security level required for an employee is determined by the sensitivity of information they will handle in their day-to-day work. During the tenure of a typical public servant, one can expect to change roles multiple times, and as such, they may need to upgrade their security clearance level. For a new public servant, assuming there are no adverse indicators in their PSS application, the high-level process for PSS is shown in Fig.~\ref{fig:PSS-process}.

After a security screening is requested by the hiring manager, the prospective employee fills out a standard application form. We note that, before the online PSS portal was introduced, this form was typically submitted in PDF format via email to a departmental security office. Some applicants even opted to print the form, complete it by hand, and send it via regular mail. Both approaches to form submission were error prone; in practice, they often led to multiple re-submissions until all errors were resolved. This led to significant processing (and thus hiring) delays. After the introduction of the online PSS system, applicants could amend their application based on errors detected by the system before submission, i.e., without having to resubmit or (e-)mail an entirely new copy.

Once the application form is accepted for processing, the necessary reference checks and background verifications take place. If no adverse indicators are found, the individual is cleared to work for the GC. Finally, before employment begins for a new employee, a security briefing takes place, which explains employees' responsibilities as public servants when handling sensitive information. For the department we examined, security briefings were conducted one-on-one with the hiring manager.


\subsection*{A.2 Handover-of-Work Model}
\label{subsec:appendix-how}

\begin{figure}
    \centering
    \includegraphics[width=3in]{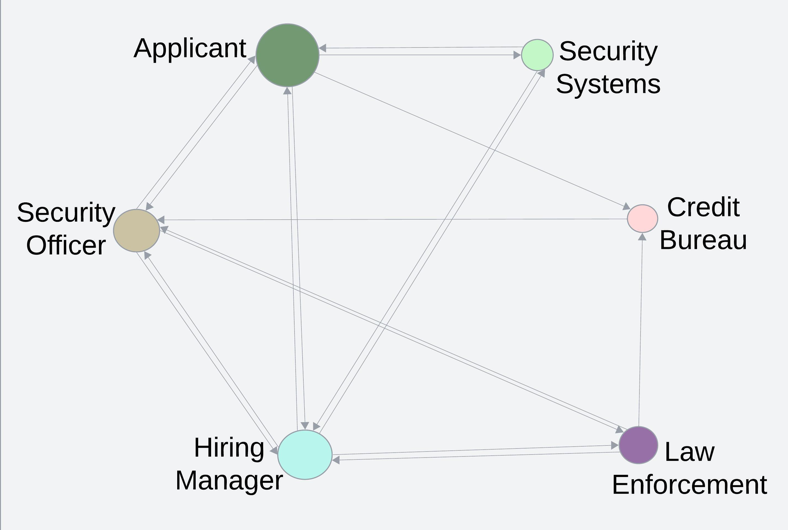}
    \caption{Handover-of-work model for the PSS process (generated with ProM).}
    \label{fig:HandoverOfWork}
\end{figure}

Figure~\ref{fig:HandoverOfWork} shows the handover-of-work model generated using ProM. Table~\ref{tab:StakeholderInvolvement} presents a more exact representation of the involvement from each stakeholder.

\begin{longtable}[]{@{}
  >{\raggedright\arraybackslash}p{(\columnwidth - 4\tabcolsep) * \real{0.3893}}
  >{\raggedright\arraybackslash}p{(\columnwidth - 4\tabcolsep) * \real{0.3521}}
  >{\raggedright\arraybackslash}p{(\columnwidth - 4\tabcolsep) * \real{0.2586}}@{}}
\caption{Stakeholder involvement in the PSS process.}
\label{tab:StakeholderInvolvement}\\
\toprule()
\begin{minipage}[b]{\linewidth}\raggedright
\textbf{Stakeholder}
\end{minipage} & \begin{minipage}[b]{\linewidth}\raggedright
\textbf{Frequency}
\end{minipage} & \begin{minipage}[b]{\linewidth}\raggedright
\textbf{Coverage}
\end{minipage} \\
\midrule()
\endhead
Applicant & 1277 & 28.4\% \\
Hiring Manager & 1159 & 25.78\% \\
Security Officer & 942 & 20.95\% \\
Law Enforcement & 547 & 12.17\% \\
Security Systems & 298 & 6.63\% \\
Credit Bureau & 273 & 6.07\% \\
\bottomrule()
\end{longtable}

\subsection*{A.3 Reasons for Application Rejections}
\label{subsec:appendix-a3}

Referring to the system's database, we ascertained the top five reasons why an application required amendments (in order of frequency from highest to lowest):

\begin{enumerate}
\def\labelenumi{\arabic{enumi}.}
\item
  The applicant omitted their middle name from the ``full given names'' field.
\item
  The applicant included their surname in the ``full given names'' field.
\item
  The applicant had an overlap in residences and resided at more than one address during the same timeframe.
\item
  The applicant uploaded a scan of an ID document that was expired at the time of application.
\item
  The applicant did not specify information about their parents.
\end{enumerate}

Other reasons for refusal included poor image quality, expiry of uploaded ID documents, the applicant's mailing address not matching their driver's license, and the applicant not including their birth name.

\subsection*{A.4 Jupyter Notebook and Event Log}
\label{subsec:appendix-artifacts}

As a companion to this case study, we have shared a sample Jupyter notebook, as well as a synthetic event log, to illustrate the use of our \texttt{logprep4pm} library. The following GitHub repository includes everything needed to start using this library, along with documentation of the functions and their associated outputs: \url{https://github.com/ProcessMining-uOttawa/logprep4pm}*
\\
\newline
* For confidentiality reasons, the artifacts provided are not from this case study, but still illustrate the general methodology that was used with a synthetic dataset.

\end{document}